\documentclass[twocolumn,showpacs,preprintnumbers,amsmath,amssymb]{revtex4}

\usepackage{graphicx,subfigure}
\usepackage{dcolumn}
\usepackage{bm}

\begin{document}

\title{Stochastic behaviour in the edge region of the SINP tokamak}
\author{Md. Nurujjaman}
\email{md.nurujjaman@saha.ac.in}
\author{Ramesh Narayanan}

\author{A.N. Sekar Iyengar}
\email{ansekar.iyengar@saha.ac.in}

\affiliation{Plasma Physics Division,
Saha Institute of Nuclear Physics,
1/AF, Bidhannagar, 
Kolkata -700064, India.}
\begin{abstract}Stochasticity is one of the most extensively researched topics in laboratory and space plasmas since it has been successful in explaining the various anomalous processes like transport, particle heating, particle loss etc. Since there is a growing need for better understanding of this nonlinear process, it has led to the development of new and more advanced data analysis techniques. In this paper we present an analysis of the floating potential fluctuations which show the existence of a stochastic multifractal process along with low dimensional chaos. This has been shown primarily by wavelet analysis, and cross checked using other nonlinear techniques. 
\end{abstract}
\maketitle

\maketitle
\section{INTRODUCTION}

Plasma is a typical complex medium exhibiting a wide variety of
nonlinear phenomena such as self oscillations, chaos, intermittency,
etc~\cite{prl:Ding,JhaKawMatt92,pramana:jaman}. The fluctuations in the edge region of magnetically confined fusion devices have also been associated to nonlinear processes like self-organization and chaotic behaviour. Interestingly it has also been shown that it is possible to have a coexistence of low dimensional chaos and stochastic behaviour~\cite{PAP-87-SigHor-PRA,01-Baketal-POP}. The goal of this paper is to present the analysis of floating potential fluctuations at the edge region of the SINP tokamak, wherein these discharges showed an enhanced emission of hard x rays signifying the loss of high energy runaway electrons~\cite{PAP-06-ramsANSI-RSI}. We have deployed several techniques, both statistical and spectral to determine the nature of stochasticity, and observe the presence of low dimensional chaos along with stochastic fractal processes similar to Bak et al~\cite{PAP-87-SigHor-PRA,01-Baketal-POP}. WTMM technique which has been successfully used in other fields,  ~\cite{MuzBacArne93,MuzBacArne91,86-Hal-PRA,arxiv:plamen,physA:Zbigniew} has been used to estimate the multifractal spectrum, and the presence of chaos, probably for the first time, in a magnetically confined plasma~\cite{PAP-03-ChaWigUze-PhysA}. We have cross checked these results with other known techniques like nonlinear analysis, probability distribution function etc.

Section~\ref{PAP-Sec-SINPTokamak} states briefly about the S.I.N.P. tokamak and in Section~\ref{PAP-Sec-ExptResults} we have presented the experimental results and and discussion and in
Section~\ref{section:conclusion} conclusion.

\section{EXPERIMENTAL SETUP: SINP TOKAMAK}\label{PAP-Sec-SINPTokamak}

The experiments were performed in the SINP Tokamak ($R_0=30 ~cm, a=7.5 ~cm$) which is a small iron core
machine having a
circular cross-section~\cite{PAP-06-ramsANSI-RSI}. In addition to the vertical magnetic field coils it also has an aluminium conducting shell
%

($R_0=30 ~cm, a=10.9 ~cm$ and thickness $=0.7 ~cm$), with  four cuts in the toroidal direction and
two in the poloidal direction respectively. Detail of the SINP tokamak will be found in Ref~\cite{PAP-06-ramsANSI-RSI}. The penetration time
of the conducting shell with cuts ($\sim 100 ~\mu s$), keeping constant the toroidal magnetic field $B_T$ at $0.8 ~T$, toroidal electric field ($E_T$) at $30.6 ~V/m$, filling pressure
$(p_{fill})$ at $0.2\pm0.05 ~mTorr$ and $a$ at $7.5 ~cm$, and $B_v$
was varied from about  $54 ~mT$ to $6.75 ~mT$. 
The edge plasma fluctuations are measured using a set of Langmuir and magnetic probes and the data were been acquired using NICOLET data acquisition system with a sampling rate of 1 MHz. 
In the present work, we report on the  analysis of the floating potential signals from an electrostatic Langmuir probe, mounted from the bottom port of the toroidal chamber, at $r=7.5 ~cm$.

\section{RESULTS AND DISCUSSION}\label{PAP-Sec-ExptResults}

\begin{figure}[htbp]
\begin{center}
\includegraphics[width=3.5 in]{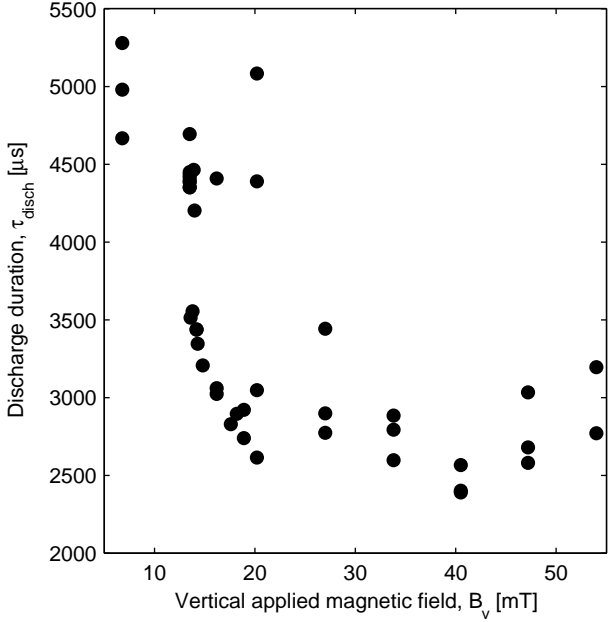}
\caption{Discharge current duration ($\tau_{disch}$) vs. $B_v$ plot.}
\label{PAP-Fig-Sch8Sch14taudischvsBv}
\end{center}
\end{figure}

\begin{figure}
\begin{center}
\includegraphics[width=3.5 in]{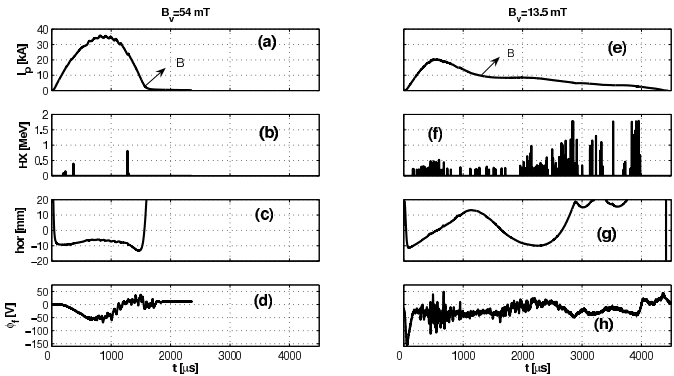}
\caption{ (a) and (e) discharge current ($I_{p}$)
evolution; (b) and (f) 3$"$ by 3$"$ NaI(Tl) limiter bremsstrahlung
bursts; (c) and (g)is the horizontal plasma positioning; (d) and (h) electrostatic probe floating potential
($\phi_f$) signals; 
for  $B_v=54 ~mT$ and $B_v=13.5 ~mT$ respectively.} \label{PAP-maxhighlowBv-IpHXLPdbdt-Sch14-1}
\end{center}
\end{figure}

An interesting behaviour of the plasma discharges was observed in the
discharge duration as the equilibrium vertical magnetic field, $B_v$ was lowered. Fig.\
\ref{PAP-Fig-Sch8Sch14taudischvsBv} shows that the discharge duration is almost constant upto $B_v\approx20 ~mT$ and then increases for $B_v<17.6 ~mT$. This extension is also clear from the discharge current at $B_v\approx13.5 ~mT$ [[Figs~\ref{PAP-maxhighlowBv-IpHXLPdbdt-Sch14-1}(e)] as compare to current duration for $54 ~mT$ [Figs~\ref{PAP-maxhighlowBv-IpHXLPdbdt-Sch14-1}(a)]. The extension
 in plasma current duration was observed after an initial fall to about half its peak value. The instant at which the
current extension is observed to begin is denoted as point B (pt. B)
[Fig~\ref{PAP-maxhighlowBv-IpHXLPdbdt-Sch14-1}]. The horizontal shift in the plasma position
($\Delta_{hor}$) has been shown in Fig.\ \ref{PAP-maxhighlowBv-IpHXLPdbdt-Sch14-1}(c) and (g) where, $+ve$
implies an outward shift.

Fig~\ref{PAP-maxhighlowBv-IpHXLPdbdt-Sch14-1}(b) shows that a few hard X-ray burst were observed at $54~mT$ when no extension in the discharge current was observed. On the other hand, the extension of the discharge after pt. B is observed to be accompanied by enhanced hard X-ray
(HX) bursts [Fig~\ref{PAP-maxhighlowBv-IpHXLPdbdt-Sch14-1}(f)]
which is indicative of loss of highly energetic particles from the
edge. A characteristic feature of the period of extension in these range
of discharges is the reduction in the electrostatic Langmuir probe
floating potential fluctuations ($\delta\phi_f$)~[Fig.\ \ref{PAP-maxhighlowBv-IpHXLPdbdt-Sch14-1}(h)]. This correlation between the reduction of fluctuations levels with the enhancement of bursts of runaway electrons, was a
motivation to study these electrostatic floating potential
fluctuations from a time-resolved statistical analysis point of
view.
\begin{figure}
\begin{center}
\includegraphics[width=3.5 in]{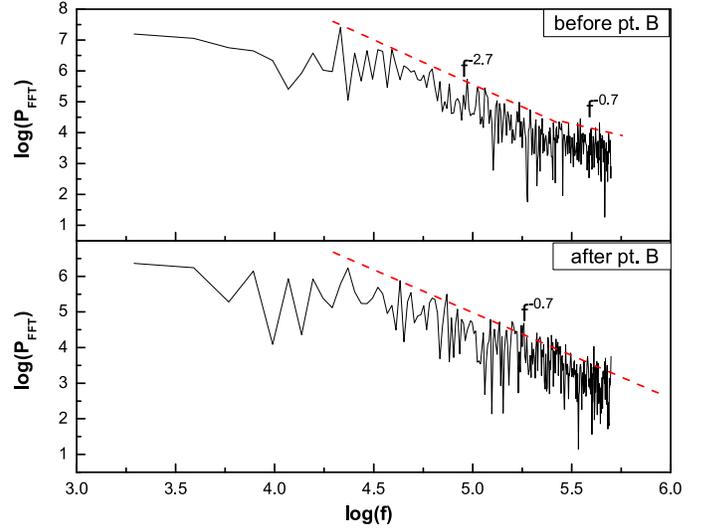}
\caption{\emph{\textbf{(color online)}} Power spectrum, in logarithmic scale (a) before pt. B, (b) after pt. B.}
\label{PAP-Fig-befaftlgFFT512pts-grph}
\end{center}
\end{figure}

It is generally accepted that if the power spectrum [$P(f)$] of a signal, obtained from Fast Fourier Transform (FFT),
decays as $P(f)\sim f^{-\beta}$ where, $\beta$ and $f$ are the the spectral index and frequency respectively, then the signal shows a stochastic fractal or self-similar behaviour\cite{PAP-87-SigHor-PRA}.  A typical plot of the
power spectrum in log-log scale is shown in Fig.\
\ref{PAP-Fig-befaftlgFFT512pts-grph}, for the discharge at $13.5 ~mT$ and it is clear that for the fluctuations before [Fig.\ \ref{PAP-Fig-befaftlgFFT512pts-grph}(a)] and after pt. B [Fig.\ \ref{PAP-Fig-befaftlgFFT512pts-grph}(a)] follow the power law behavior, which indicates the presence of stochastic fractal processes. As the power spectrum cannot extract the information regarding the time-frequency simultaneously, the presence of sharp transitions and small scale features contained in the signal, we introduce more advanced techniques like wavelet analysis etc.

\textbf{Wavelet analysis:} 

 Wavelet analysis\cite{book:mallat,PAP-92-Daubechies-BK, PAP-95-Holschneider-BK}, provides a
way of analyzing the local behaviour of functions and correct
 characterization of time series in the presence of non-stationarity like global or
local trends or biases. One of the main aspects of the Wavelet analysis is
of great advantage is the ability to reveal the hierarchy of
(singular) features, including the scaling behaviour
\cite{PAP-95-ArnBacMuz-PA}.
\begin{figure}
\begin{center}
\includegraphics[width=3.5 in]{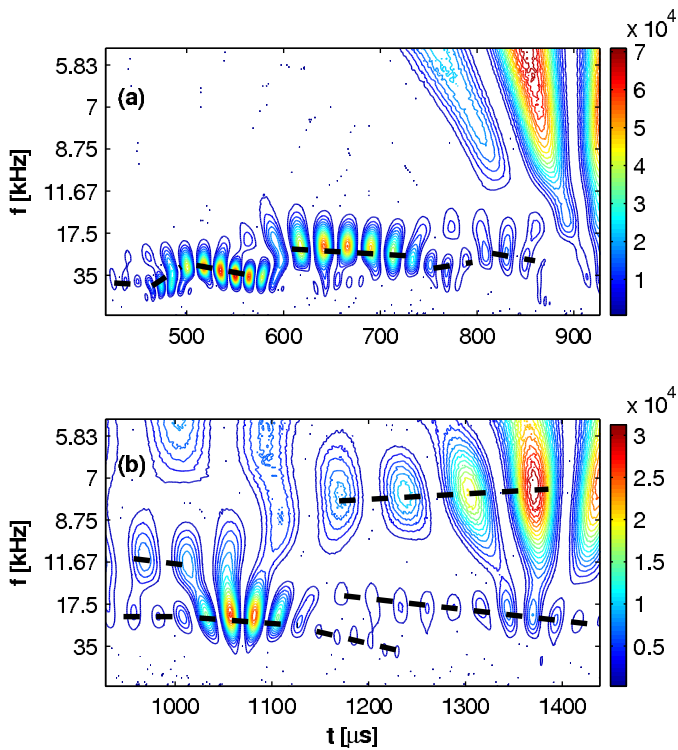}
\caption{\emph{\textbf{(color online)}}Time-frequency contour plot of power spectrum using wavelet transform, intensity of the power have given by color bar: (a)
before pt. B, (b) after pt. B. {\emph{Black dashed lines}} show the
horizontal ridges connecting maxima values in contour plot.}
\label{PAP-Fig-lowBv-CWTanalysis-befaftptB-Sch14}
\end{center}
\end{figure}

\begin{figure}
\begin{center}
\includegraphics[width=3.5 in]{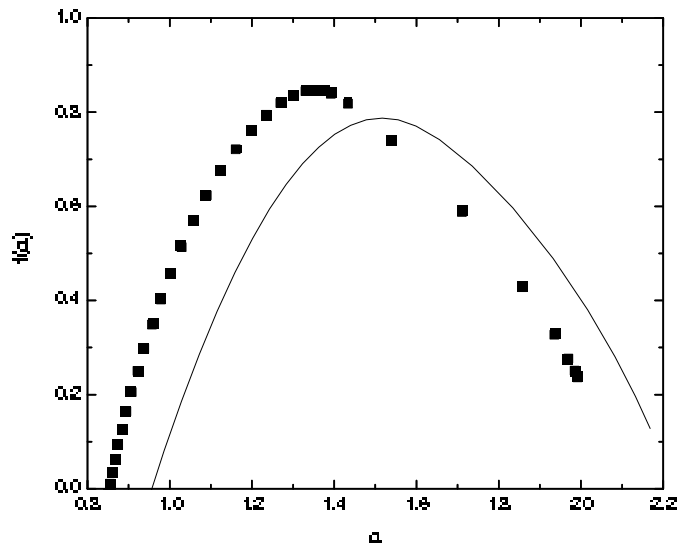}
\caption{Typical multifractal singularity spectrum 
 before (solid line) and after pt. B(dot dot line)  at 13.5 mT.}
\label{PAP-Fig-befaft-Dhsel-grph}
\end{center}
\end{figure}

The wavelet transform of a
function $\phi(t)$ is then given by:
\begin{equation}
\label{eqn:wt}
W_\Psi(s,\tau)=\int \phi(t)\Psi_s(t-\tau)dt
\end{equation}
where, $\phi(t)$ is the signal and $\Psi(t)$ is an oscillating functions that decays
rapidly with time and are termed as wavelets. s and $\tau$ are the scale and time respectively. Fig.\ \ref{PAP-Fig-lowBv-CWTanalysis-befaftptB-Sch14} represents the time-frequency contour plot of the power spectrum ($\mid W_{\Psi}$(s,$\tau$)$\mid^2$) obtained from the wavelet analysis [Eqn.~\ref{eqn:wt}], for the discharge
at $13.5 ~mT$. For simplicity the scales in
the y-axis have been converted to psuedo-frequency ($F_s$) which has been estimated from the relation $F_s=\frac{F_c}{s\Delta}$~\cite{web:scal2frq}, where $F_c$ is center frequency of the analyzing wavelet and $\Delta$ is the sampling period of the signal. The presence of chaos or periodicity can be studied using ridge plots obtained from the wavelet transform spectrum, which has been discussed in detail by Chandre et al~\cite{PAP-03-ChaWigUze-PhysA}. 
Fig.\ \ref{PAP-Fig-lowBv-CWTanalysis-befaftptB-Sch14}(a) shows the typical ridge plot at $13.5~mT$
 before pt B. It shows that the most of the power is concentrated almost at a constant time scale a. 
The connected horizontal ridges  suggest that the
electrostatic floating potential signals are quasi periodic with
resonance transitions occurring at regular intervals~\cite{PAP-03-ChaWigUze-PhysA}.
For the signal after pt. B, the power is concentrated in two or three modes simultaneously
 at any given instant of time [\ref{PAP-Fig-lowBv-CWTanalysis-befaftptB-Sch14}(b)] indicating the presence of chaos~\cite{PAP-03-ChaWigUze-PhysA}.

The singularity spectrum, $f(\alpha)$ vs. $\alpha$, where $f(\alpha)$ is the distribution of the singularity strength $\alpha$, has been estimated using Wavelet Transform Modulus Maxima (WTMM) method~\cite{MuzBacArne93,MuzBacArne91,86-Hal-PRA} 
using the following canonical equations,

\begin{eqnarray}\label{fracdimWTMMcanonicalmethoda}
\noindent \alpha(q)=\lim_{s\rightarrow 0} \frac{1}{\ln s}
\sum_{\{t_i(s)\}_i}&&
\hat{W}_\Psi(q;s,t_i(s))\times\nonumber\\&&\ln|W_{\Psi}(s,t_i(s))|\nonumber\\
\\\label{fracdimWTMMcanonicalmethodb} \noindent f(\alpha(q))=\lim_{s\rightarrow 0}
\frac{1}{\ln s} \sum_{\{t_i(s)\}_i}&&
\hat{W}_\Psi(q;s,t_i(s))\times\nonumber\\&&\ln|\hat{W}_\Psi(q;s,t_i(s))|\nonumber\\
\end{eqnarray}
where,
$\hat{W}_\Psi(q;s,t_i(s))=\frac{{|W_{\Psi}(s,t_i(s))|}^q}{\sum_{\{t_i(s)\}_i}{|W_{\Psi}(s,t_i(s))|}^q}$

The singularity spectrum for before and after pt B, for discharge at $13.5~mT$ is shown in Fig.\
\ref{PAP-Fig-befaft-Dhsel-grph}. The spectrum seems to be slightly asymmetric before pt. B, whereas it is almost symmetric after pt. B. The symmetry gives one an indication of multiplicative process~\cite{AntDevGarLuck2001} and hence the  fluctuations in the extended phase is associated to some avalanche phenomena.
\begin{figure}
\begin{center}
\includegraphics[width=3.5 in]{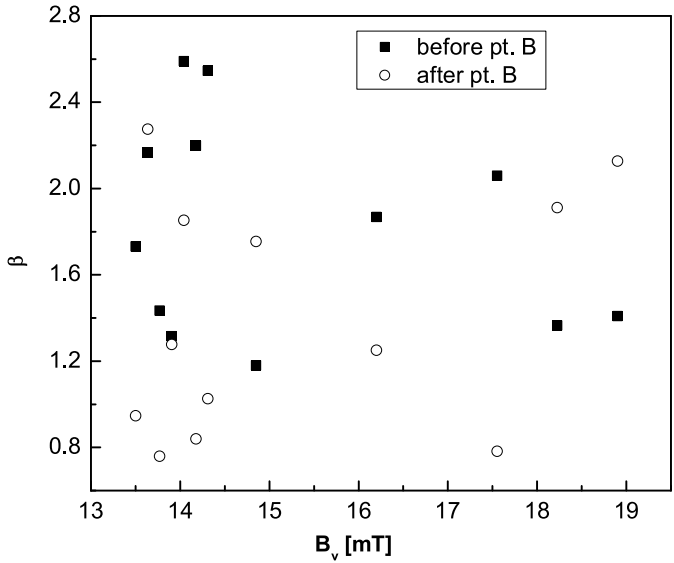}
\caption{Degree of multifractality ($\beta$) 
pts before [\emph{solid square}] and after pt. B
[\emph{open circle}] for the various discharges of $B_v$.}
\label{PAP-Fig-befaftbeta-grph}
\end{center}
\end{figure}
\begin{figure}
\begin{center}
\includegraphics[width=3.5 in]{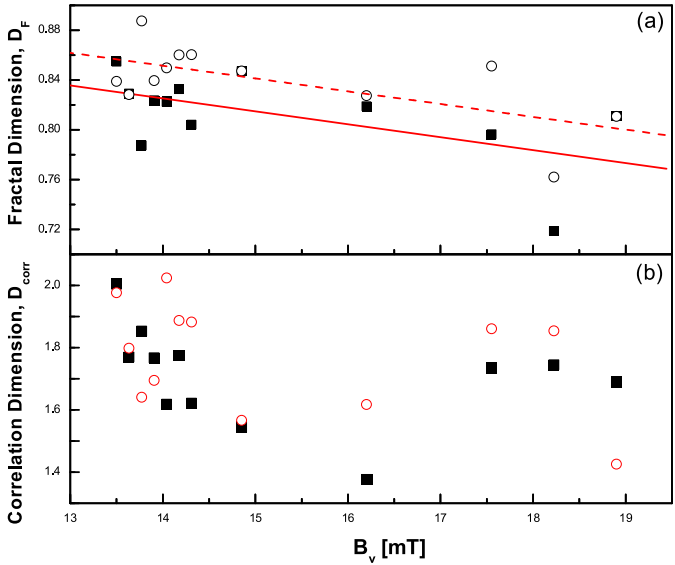}
\caption{\emph{\textbf{(color online)}}(a) $D_F$ before
[\emph{black, solid square}] and after pt. B [\emph{black, open
circle}]  and (b) $D_{corr}$ before [\emph{black, solid square}] and after pt. B [\emph{orange, open circle}] for the various  $B_v$. } \label{PAP-Fig-DF-Dcorr-befaft-grph}
\end{center}
\end{figure}

The characteristic of the signals can also be described by degree
of the multifractality ($\beta$) which is defined by the
difference between the maximum singularity strength
($\alpha_{max}$) and the minimum singularity strength
($\alpha_{min}$)~\cite{pre:chhabra,01-SilHu-PRE,04-BudKikUesTak-NF,LectNotesPhys:jaman}.
Fig.\ \ref{PAP-Fig-befaftbeta-grph} shows the range of $\beta$ for $B_v<15
~mT$ is $0.6-1.2$ for datasets after pt. B and it is
$1.2-2.7$ for datasets before pt. B. The decreasing trend in $\beta$ for the extended phases indicates that the system has a tendency to go towards a stochastic state.

We estimated the fractal dimension ($D_F$) and correlation
dimension ($D_{corr}\equiv D(q=2)=2\alpha-f(\alpha)$) from the
singularity spectrum~\cite{01-Baketal-POP}. Fig.\
\ref{PAP-Fig-DF-Dcorr-befaft-grph}(a) and Fig.\
\ref{PAP-Fig-DF-Dcorr-befaft-grph}(b) show that in the extended discharge $D_F$ and $D_{corr}$ are in the range of $0.86-0.84$ and $1.8-2.0$ respectively, indicating the presence of complex nature in the signal.

\begin{figure}
\begin{center}
\includegraphics[width=3.5 in]{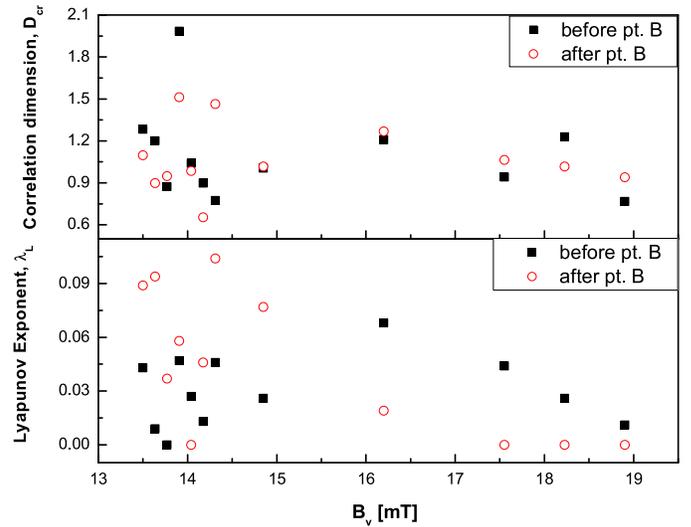}
\caption{\emph{\textbf{(color online)}}(a) $D_{cr}$  before
[\emph{black, solid square}] and after pt. B [\emph{orange, open
circle}]  and (b) $\lambda_L$  before [\emph{black, solid square}] and after pt. B [\emph{orange, open
circle}] for the various discharges of $B_v$.} \label{PAP-Fig-Dcorr-Lyap-befaft-grph}
\end{center}
\end{figure}

A crosscheck of the above results for the presence of chaos or complexity, can be made by estimating the correlation
dimension ($D_{cr}$) and Lyapunov exponent ($\lambda_L$). $D_{cr}$ and $\lambda_L$ have been estimated using the Grassberger-Procaccia techniques~\cite{physrevlett:grassberger,physrevA:grassberger} and the Wolf algorithm
\cite{physicaD:Wolf,book:sprott} respectively. $D_{cr}$ and $\lambda_L$
before and after pt. B have been presented in Fig.\
\ref{PAP-Fig-Dcorr-Lyap-befaft-grph}(a) and \ref{PAP-Fig-Dcorr-Lyap-befaft-grph}(b) respectively. From Fig.~\ref{PAP-Fig-Dcorr-Lyap-befaft-grph}(a), it is clear that correlation dimension obtained using multifractal analysis and Grassberger-Procaccia techniques are of same order. Fig.\ \ref{PAP-Fig-Dcorr-Lyap-befaft-grph}(b) shows $\lambda_L$ is more positive for $B_V<16~mT$ after pt. B indicating chaos. Though we have estimated these exponent using insufficient less of data points, the results agrees well with wavelet analysis.

 \begin{figure}
 \begin{center}
    \subfigure[]{\includegraphics[width=2.5 in]{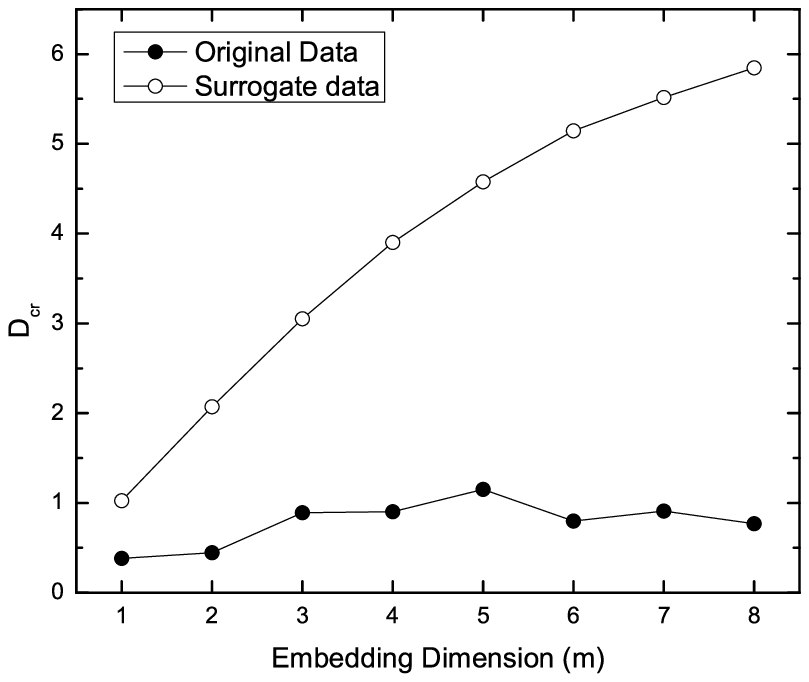}}
    {
      \label{PAP-Fig-shuffle-Dcorr-aft-grph1}
    }
    \subfigure[]{\includegraphics[width=2.75 in]{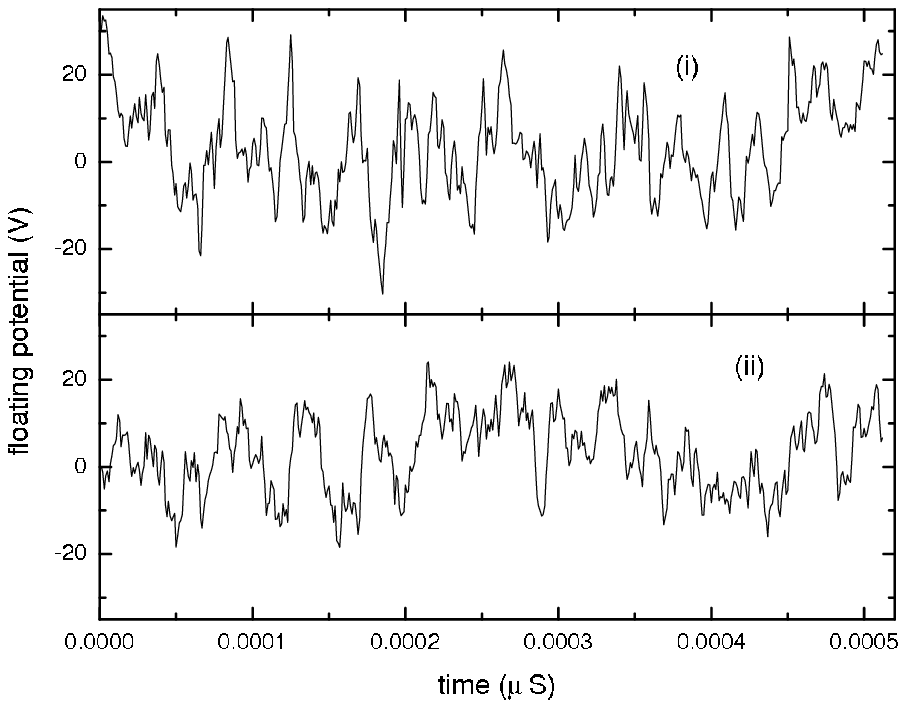}}
    {
      \label{PAP-Fig-shuffle-Dcorr-aft-grph2}
    }
  \caption{(a). Correlation dimension
($D_{cr}$) against embedding dimension for original \emph{solid
circle} and surrogate \emph{open circle} dataset. The dataset used
is of the data after pt. B in
Fig.~\ref{PAP-maxhighlowBv-IpHXLPdbdt-Sch14-1}(i). (b). (i) Original data and (ii) corresponding phase shuffled surrogate data.}
  \label{PAP-Fig-shuffle-Dcorr-aft-grph}
  \end{center}
  \end{figure}

In order to validate our nonlinear analysis we did a
surrogate analysis of the extended discharge regime of Fig.\
\ref{PAP-maxhighlowBv-IpHXLPdbdt-Sch14-1}(i). The surrogate data
has been generated by Phase Shuffled surrogate method, in which
the phases are randomized by shuffling the Fourier
phases~\cite{chaos:dori,physicaD:theiler,Bifur:Nakamura}, and
hence  the power spectrum (linear structure) is preserved, but the
nonlinear structures are destroyed~\cite{Bifur:Nakamura}. $D_{cr}$
has been estimated for both the original [Fig~\ref{PAP-Fig-shuffle-Dcorr-aft-grph}(b)i] and the corresponding
surrogate data [Fig~\ref{PAP-Fig-shuffle-Dcorr-aft-grph}(b)ii], shown in Fig~\ref{PAP-Fig-shuffle-Dcorr-aft-grph}(a)
by \emph{solid circle} and \emph{open circle} respectively. The
$D_{cr}$ for the original data saturates at higher m, whereas in
the case of the surrogate data one finds $D_{cr}$ keeps on
increasing with m. Hence the estimated $D_{cr}$ and $\lambda_L$ are
from nonlinear effects in the system.

\begin{figure}
\begin{center}
\begin{center}
\includegraphics[width=3.5 in]{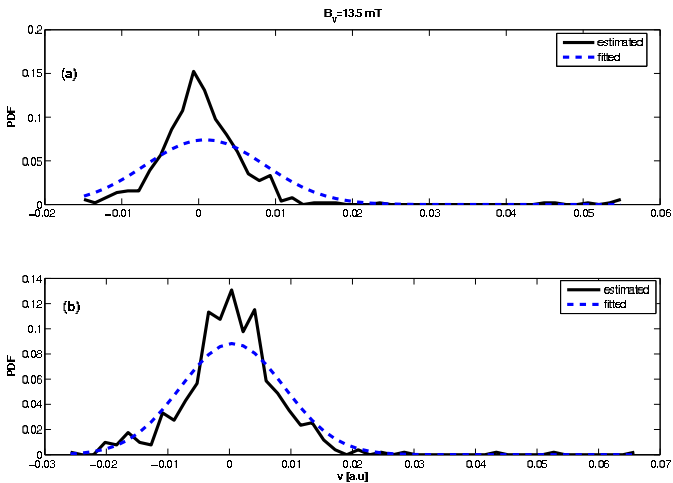}
\caption{(color online) PDF of edge electrostatic floating potentials: before
[(a)] and after pt. B [(b))] at $B_v13.5 ~mT$. Corresponding gaussian fitting is shown by dashed line.}
\label{PAP-Fig-Sch14_BLP1_PDF_twoBv}
\end{center}
\end{center}
\end{figure}

\begin{figure}
\begin{center}
\includegraphics[width=3.5 in]{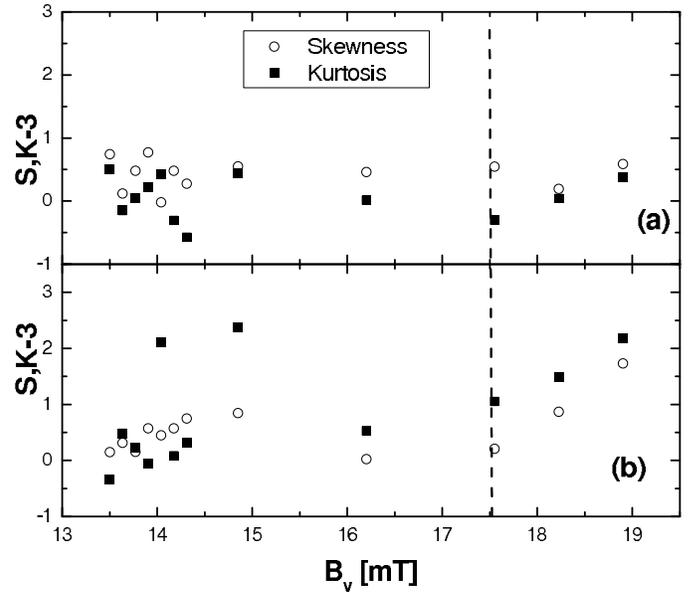}
\caption{Skewness (S) [black open circle] and (K-3)
[black, solid square] (a) before and
(b) after pt. B for various values of $B_v$. The dotted vertical line demarcates either side of $B_v=17.6 ~mT$}
\label{PAP-Fig-bef-skwkrt-grph}
\end{center}
\end{figure}

Probabilistic descriptions such as {\underline{P}}robability {\underline{D}}istribution
{\underline{F}}unction  [PDF] are at the heart of the characterization of turbulence~\cite{book:frisch}. 
Fig.~\ref{PAP-Fig-Sch14_BLP1_PDF_twoBv}(a) and \ref{PAP-Fig-Sch14_BLP1_PDF_twoBv}(b)show the PDF at $13.5 ~mT$ 
before and after pt. B respectively. The corresponding gaussian fitting is shown by dashed line. Both plots show that the PDFs are non-Gaussian in nature. Skewness (S) and Kurtosis (K) which are measure of nongaussianity are shown in  Fig.\ \ref{PAP-Fig-bef-skwkrt-grph}, for the extended discharge which also indicate deviation from gaussianity~\cite{book:frisch}. From above analysis it is clear that during the extended discharge 
phase neither the system is purely stochastic in nature nor chaotic, rather a mixture of both is present.
  
\begin{figure}
\begin{center}
\includegraphics[width=3.5 in]{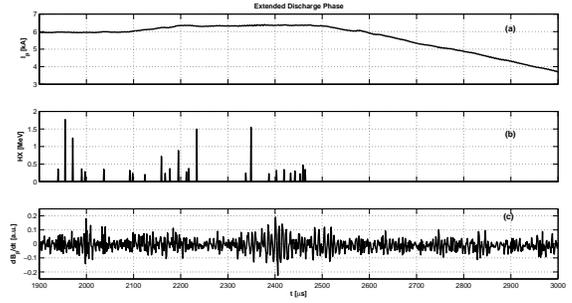}
\caption{Plot of discharge characteristics depicting (a) Plasma
Current, $I_p$ in kA, (b) HX bursts in MeV, (c) dB/dt signals.}
\label{PAP-Fig-Sch14-IpHXdBdt}
\end{center}
\end{figure}

The observations of the enhanced energy levels in the HX bursts in
the extended phase could be a result of loss of the
high energy particles which are probably generated in this phase. The observations of HX bursts can be correlated to some growing modes in the dB/dt signals, as
during the time instants of $1920 ~\mu s-2050 ~\mu s$,  $2090 ~\mu
s-2230 ~\mu s$ and  $2320 ~\mu s-2500 ~\mu s$ (Fig.\
\ref{PAP-Fig-Sch14-IpHXdBdt}). Subsequently, one could infer that
some instability could be triggering the deconfinement of the
particles, which are thereafter lost, through some stochastic
process at the edge.

To cross-check whether the discharge could sustain any such
beam-plasma interactions, we considered the conditions that need
to be satisfied~\cite{PAP-02-Plyusnin-EPSConf}:
$\omega_{ce}>\omega_{pe}$,
 $\nu_{eff}>\nu_{e}$, and
$v_{beam}>3v_{cr}\biggl{(}\frac{\omega_{ce}}{\omega_{pe}}\biggr{)}$,
where, $\omega_{ce}$, $\omega_{pe}$, $\nu_e$, $v_{beam}$ and  $v_{cr}$ are the electron cyclotron frequency, electron plasma frequency, electron collision frequency, beam velocity and the critical velocity respectively and
$\nu_{eff}=\sqrt{\pi}\omega_{pe}\biggl{(}\frac{\omega_{pe}}{\omega_{ce}}\biggr{)}\epsilon^{-3/2}\lambda_r$,
$\lambda_r$ being the primary runaway flux generation factor and
$\epsilon=E_0/E_{cr}$, $E_{cr}$ being the critical electric field
for runaway generation and $E_0=V_{loop}/2\pi R_0$. $V_{loop}$ is
the loop voltage. Using the experimental results, for beam energy $\approx1~MeV$, $n_e\approx6\times10^{18}~m^{-3}$, $B_T\approx0.8~T$, $V_{loop}\approx40~V$ and $Z_{eff}=1$, we have  $\omega_{ce}\approx 140.5$ GHz and
$\omega_{pe}\leq 100$ GHz and $\nu_{e}=300$ kHz and $\nu_{eff}=1.2$ MHz. Hence first and second condition are satisfied in the extended discharge phase. For the same experimental conditions, $3v_{cr}[\omega_{ce}/\omega_{pe}]\approx 3\times10^7~m/s$ before pt B which corresponds to a
beam energy of 2keV and  after pt B, $3v_{cr}[\omega_{ce}/\omega_{pe}]\approx 10^8~m/s$ (corresponding beam energy is 30 keV). Thus in the current extension phase the more energetic electrons will satisfy the third condition. Hence it could imply that the loss of runway electrons observed in the extended phase could be a result of the
participation of the higher energy electrons in beam-plasma instabilities within the plasma column. The edge stochastic behaviour possibly leads to the ejection of these runaway electrons.

\section{CONCLUSION}
\label{section:conclusion}
In the extended discharges of the SINP tokamak, where enhanced HX emission were observed, 
we have shown the presence of combination of both of stochasticity and low dimensional 
chaos using various techniques like wavelet analysis and other 
nonlinear techniques like the estimation of correlation dimension, Lyapunov exponent and PDF.
One still needs to look into other edge fluctuation behaviour,
such as the magnetic, density and temperature fluctuations, in
order to understand the role of the stochastic behaviour with the
discharge extension, using the wavelet transform especially the WTMM method.

\section*{ACKNOWLEDGEMENTS}
We would like to thank Prof. B. Sinha, Director, SINP for his
support in carrying out this work. We also thank members of
Plasma Physics Division, SINP for their help during the experiments. RN would
like to acknowledge useful discussions with Prof. V.P. Budaev and Prof. M. Rajkovic and  Prof. K.H. Finken and organizers  for providing financial support to attend Fusion Plasmas-2007 Workshop, Julich Germany.
MN  acknowledges discussions with Prof. J. C. Sprott on nonlinear analysis techniques.

\end{document}